%% file: interspeech2022.tex
\documentclass[a4paper]{article}

\usepackage{INTERSPEECH2021}
\usepackage[printonlyused, nolist]{acronym}
\graphicspath{{../figs/}}
\usepackage{xcolor}
\usepackage{stmaryrd}
\usepackage{url}
\usepackage{subcaption}

\title{SAMbA: Speech enhancement with Asynchronous ad-hoc Microphone Arrays}
\name{Nicolas Furnon$^1$, Romain Serizel$^1$, Slim Essid$^2$, Irina Illina$^1$}
\address{
  $^1$Université de Lorraine, CNRS, Inria, Loria, F-54000 Nancy, France\\
  $^2$LTCI, Télécom Paris, Institut Polytechnique de Paris, Palaiseau, France}
\email{$^1$\{firstname.lastname\}@loria.fr, $^2$slim.essid@telecom-paristech.fr}

\begin{document}
\input{acronyms.tex}

\maketitle
\begin{abstract}
Speech enhancement in ad-hoc microphone arrays is often hindered by the 
asynchronization of the devices composing the microphone array. Asynchronization
comes from sampling time offset and sampling rate offset which inevitably occur
when the microphones are embedded in different hardware components.
In this paper, we propose a deep neural network (DNN)-based speech enhancement solution that
is suited for applications in ad-hoc microphone arrays
because it is distributed and copes with asynchronization. We show that
asynchronization has a limited impact on the spatial filtering and mostly
affects the performance of the DNNs. Instead of resynchronising the signals,
which requires costly processing steps, we use an attention mechanism which
makes the DNNs, thus our whole pipeline, robust to asynchronization. We also
show that the attention mechanism leads to an estimation of the sampling time offset in an unsupervised manner.

\end{abstract}

\section{Introduction}
Due to their increased number of microphones, their spatial coverage and their flexibility of use, ad-hoc microphone arrays offer great potential to speech enhancement. This potential however may be limited by a series of challenges.
One of the main challenges is the need for a distributed strategy which does not rely on a fusion center as most of classic beamformers do. Distributed algorithms have been proposed for speech enhancement in ad-hoc microphone arrays \cite{Bertrand2010a, OConnor2014, Markovich-Golan2015, Sherson2016} and recently, a \ac{dnn}-based distributed solution has also been introduced to combine the increased modelling capacity of \acp{dnn} with the flexibility of use of ad-hoc microphone arrays~\cite{Furnon2021}.
Besides, because the microphones embedded in different devices do not share the same hardware, they are acquired at different \acfp{sr} (even if the nominal \ac{sr} is the same), causing a \ac{sro}, and triggered at different starting times, causing a \ac{sto}. These phenomena cause asynchronization  \cite{Schmalenstroeer2015, Ceolini2020}. Asynchronization can have a negative impact on speech enhancement \cite{Lienhart2003}, especially on solutions relying on an accurate estimation of the direction-of-arrival, like the minimum variance distortionless response beamformer \cite{Cherkassky2015, Zeng2015, Schmalenstroeer2018}. It is yet interesting to note that some works showed that asynchronization could be tolerated in speech enhancement tasks without any attempt to resample the signals \cite{Chiba2014, Corey2018}.

Solutions to asynchronization can be broadly classified into two categories. In the first category, specific signals are sent among the nodes. These signals can be either calibration signals \cite{Lienhart2003, Wehr2004} or time stamps \cite{Schenato2011, Schmalenstroeer2015, Ceolini2020}. The second category gathers so-called \textit{blind} approaches, because no signals are exchanged but the ones captured by the microphones of the nodes. Out of these observations, the \ac{sro} and \ac{sto} can be estimated and compensated for based on the coherence \cite{Markovich2012b, Schmalenstroeer2017}, the correlation \cite{Wang2016} or the cross-correlation \cite{Miyabe2013, Cherkassky2014, Schmalenstroeer2017, Chinaev2021} between signals of different nodes. These solutions proved to be efficient, but they suffer from two limitations. The first one is that they require extra computing steps, which might overload the small devices of ad-hoc microphone arrays and add some latency to the processing. The second limitation is that none of these works studies the impact of asynchronization on \acp{dnn} although \acp{dnn} take a more and more important place in speech enhancement.

In this paper, we propose to study the impact of asynchronization on a speech enhancement solution for ad-hoc microphone arrays based on \acp{dnn}. We show that the impact of \ac{sro} and \ac{sto} on spatial filtering is limited, but that their impact on the \ac{dnn} performance is not negligible. To cope with this, instead of resampling the signals, we decide to use an attention mechanism which implicitly realigns the input signals of the \ac{dnn}. This avoids an explicit search of the asynchronization parameters.

This paper is organized as follows. In Section~\ref{sec:problem} we describe the problem, the notations used throughout the paper and we introduce our speech enhancement system. The experimental setup is described in Section~\ref{sec:setup}. In Section~\ref{sec:async} we analyse the impact of \ac{sto} and \ac{sro} on our speech enhancement system. In Section~\ref{sec:solution}, we introduce a solution to cope with the asynchronization effects on the \ac{dnn} performance in our system. Lastly, Section~\ref{sec:conclusion} concludes this paper.

\section{Problem formulation}\label{sec:problem}

\subsection{Notations}
In the following, signals are considered in the \ac{stft} domain, where time and frequency indices are dropped for the sake of conciseness. Bold lowercase letters represent vectors. Bold uppercase letters represent matrices. Regular lowercase represent scalars.
We consider an ad-hoc microphone array of $K$ nodes of $M_k$ microphones each. The $m$-th microphone of the $k$-th node records a noisy mixture $y_{k, m}~=~s_{k, m} + n_{k, m}$ according to an additive noise model, where $s_{k, m}$ and $n_{k, m}$ are respectively the target speech and the noise components recorded by the microphone. The signals recorded by node $k$ are stacked in a vector $\mathbf{y}_k~=~[y_{k, 1}, \cdots , y_{k, M_k}]^T$ .

The \ac{sro} of a node $k$ relatively to a reference node will be denoted by $\epsilon_k$. 
The \ac{sto} of a node $k$ relatively to a reference node will be denoted by $\tau_k$. 

\subsection{Distributed speech enhancement in ad-hoc microphone arrays}
\label{subsec:tango}
\begin{figure}
	\centering
	\includegraphics[width=\linewidth, trim=0cm 0cm 1cm 0cm, clip]{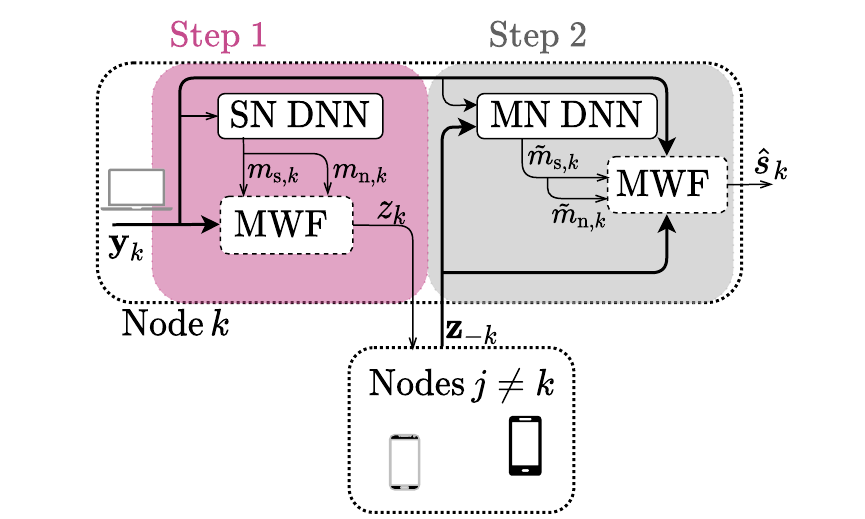}
	\caption{Graphical representation of our distributed speech enhancement solution. Bold arrows represent multichannel signals, simple arrows represent single-channel signals.}
	\label{fig:tango}
\end{figure}
In a previous work, we introduced a distributed speech enhancement system for ad-hoc microphone arrays, called Tango \cite{Furnon2021}. It processes in two steps, highlighted in Figure~\ref{fig:tango}. In the first step,  at each node $k$, a \ac{mwf} $\mathbf{w}_{kk}$ is applied on the local signals $\mathbf{y}_k$. To do this, a \ac{sndnn} is used to predict a \ac{tf} mask $m_k$ out of the reference signal $y_{k, 1}$. The \ac{tf} mask is used to compute the spatial covariance matrices of the speech and noise required by the spatial filter.

Filtering the mixture with this beamformer yields a so-called compressed signal 
$
z_{k} =  \mathbf{w}_{kk}^H\mathbf{y}_k\,.
$
The compressed signals are exchanged among nodes, so node $k$ receives $K - 1$ compressed signals $\mathbf{z}_{-k}$: \hfil\linebreak
$\mathbf{z}_{-k}~=~[z_1, ..., z_{k-1}, z_{k+1}, ..., z_K]^T$. In the second step, a global \ac{mwf} $\mathbf{w}_{k}$ is applied on $\tilde{\mathbf{y}}_k = \left[ \mathbf{y}_k^T,~\mathbf{z}_{-k}^T\right]^T$. The compressed signals $\mathbf{z}_{-k}$ are used for the spatial filtering operation, but they are also fed to a \ac{mndnn} to predict the \ac{tf} mask $\tilde{m}_k$ required by the spatial filter.
The spatial filters at both filtering steps are computed following the \mbox{rank-1} \ac{gevd} of the covariance matrices of the mixture and of the noise proposed by Serizel et al.\ \cite{Serizel2014}. 

We showed that this algorithm could efficiently process the spatial information conveyed by the compressed signals and outperforms an oracle \ac{vad}-based \ac{mwf}~\cite{Furnon2020}. We also showed that it performs comparatively well to FaSNet \cite{Luo2019}, while allowing for a trade-off between noise reduction and speech distortion, and relying on a much simpler \ac{dnn} architecture \cite{Furnon2021}. For these reasons, we continue using this system for the current work.

\section{Experimental setup}\label{sec:setup}
\subsection{Signal setup and \ac{dnn} settings}
All the signals are sampled at 16~kHz and last between 5~s and 10~s. The \ac{stft} is computed with a Hann window of 32~ms with an overlap of 16~ms. The \ac{crnn} architecture is composed of three convolutional layers followed by a recurrent layer and a fully-connected layer. The convolutional layers have 32, 64 and 64 filters, with kernel size $3 \times 3$ and stride $1 \times 1$. Each convolutional layer is followed by a batch normalisation and a maximum-pooling layer of kernel size $4 \times 1$ so that no pooling is applied over the time axis. The recurrent layer is a 256-unit GRU. The fully-connected layer has 257 units with a sigmoid activation function. The input of the model are the magnitudes of the \ac{stft} windows of 21 consecutive frames and the ground truth labels are the corresponding frames of the ideal ratio mask. At test time, only the middle frame of the predicted window is considered to estimate the mask, so sliding windows of the input are fed to the \ac{dnn}. The mask of the whole signal is predicted before being used to enhance the speech in a batch mode.

\subsection{Training and evaluation data}
\label{subsec:setup_data}
The data used to train and evaluate our systems is extracted from the \textsc{Disco} dataset.\footnote{\url{https://github.com/nfurnon/disco/tree/master/dataset_generation}} Room impulse responses in shoebox-shaped rooms are simulated. The rooms have a length, width and height randomly picked in the ranges $\llbracket 3; 8\rrbracket$~m, $\llbracket 3; 5\rrbracket$~m and $\llbracket 2; 3\rrbracket$~m respectively. 2 sources, one target source and one noise source, are randomly laid in the room. 4 nodes of 4 microphones each are randomly laid in the room and record the scene. The only constraint is that the sources and the microphones should not be closer than 50~cm from each other and from the walls.

In our experiments, the effects of \ac{sro} and \ac{sto} are considered separately. Their joint impact is left for future work.
To simulate asynchronization, in each simulated configuration, one node $k$ among the four is chosen as the reference node. Its \ac{sr} (resp. sampling start) is left unchanged, this is why we have $\epsilon_k=0$ (resp. $\tau_k=0$) for this node.
The \ac{sro} is simulated by resampling the signals of nodes $j\ne k$ at various sampling frequencies.
The \ac{sto} is simulated by padding zeros at the beginning of the signals of nodes $j \ne k$.
Because of the symmetry of the \ac{sro} and \ac{sto} effects, only positive \acp{sro} and \acp{sto} will be considered.
For each node $j\ne k$, the \ac{sro} $\epsilon_j$ is randomly taken between 0~parts per million (ppm) and a maximum value $\text{SRO}_{\text{max}}$. 6 different values of $\text{SRO}_{\text{max}}$ are considered, leading to 6 evaluation conditions.
The \ac{sto} is randomly taken between 0~ms and a maximum value $\text{STO}_{\text{max}}$. 5 different values of $\text{SRO}_{\text{max}}$ are considered, leading to 5 evaluation conditions.
Because the signals of one node share the same hardware and software implementation, we assume that they are synchronized. As a consequence, asynchronization can only affect the second filtering step of Tango.

\section{Impact of asynchronization on \ac{dnn}-based speech enhancement}
\label{sec:async}
In this section, the system described in section \ref{subsec:tango} is evaluated on the data described in section \ref{subsec:setup_data}. However, the \acp{mndnn} of the second filtering step are trained with synchronized data: at train time, the compressed signals received by a given node are perfectly synchronized with the mixtures recorded by the receiving node.

\begin{figure}[h!]
	\centering
	\subfloat[\ac{sro} impact]{
		\includegraphics[width=.65\linewidth]{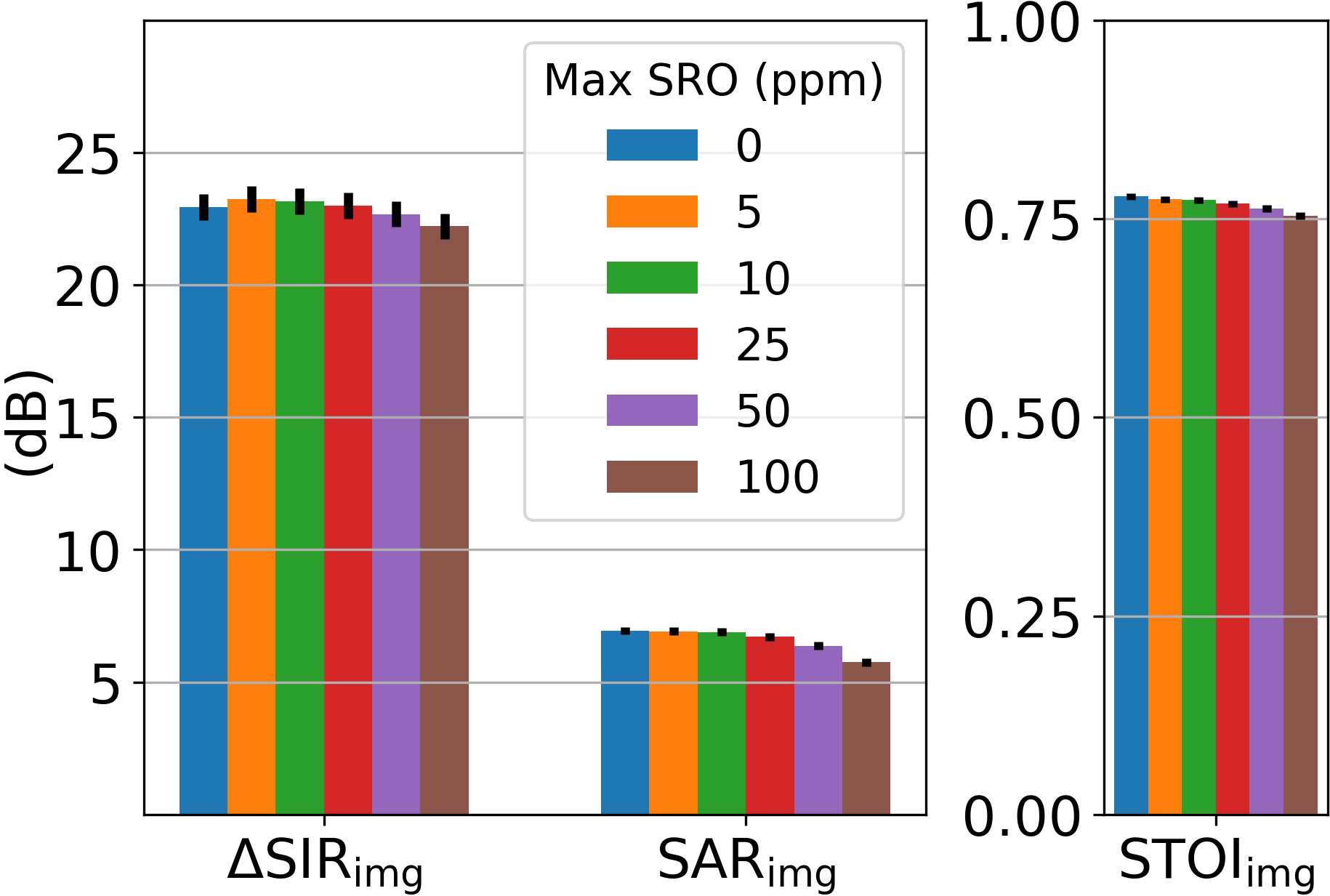}%
		\label{subfig:async_impact_sro}}
	\hfil
	\subfloat[\ac{sto} impact]{
		\includegraphics[width=.65\linewidth]{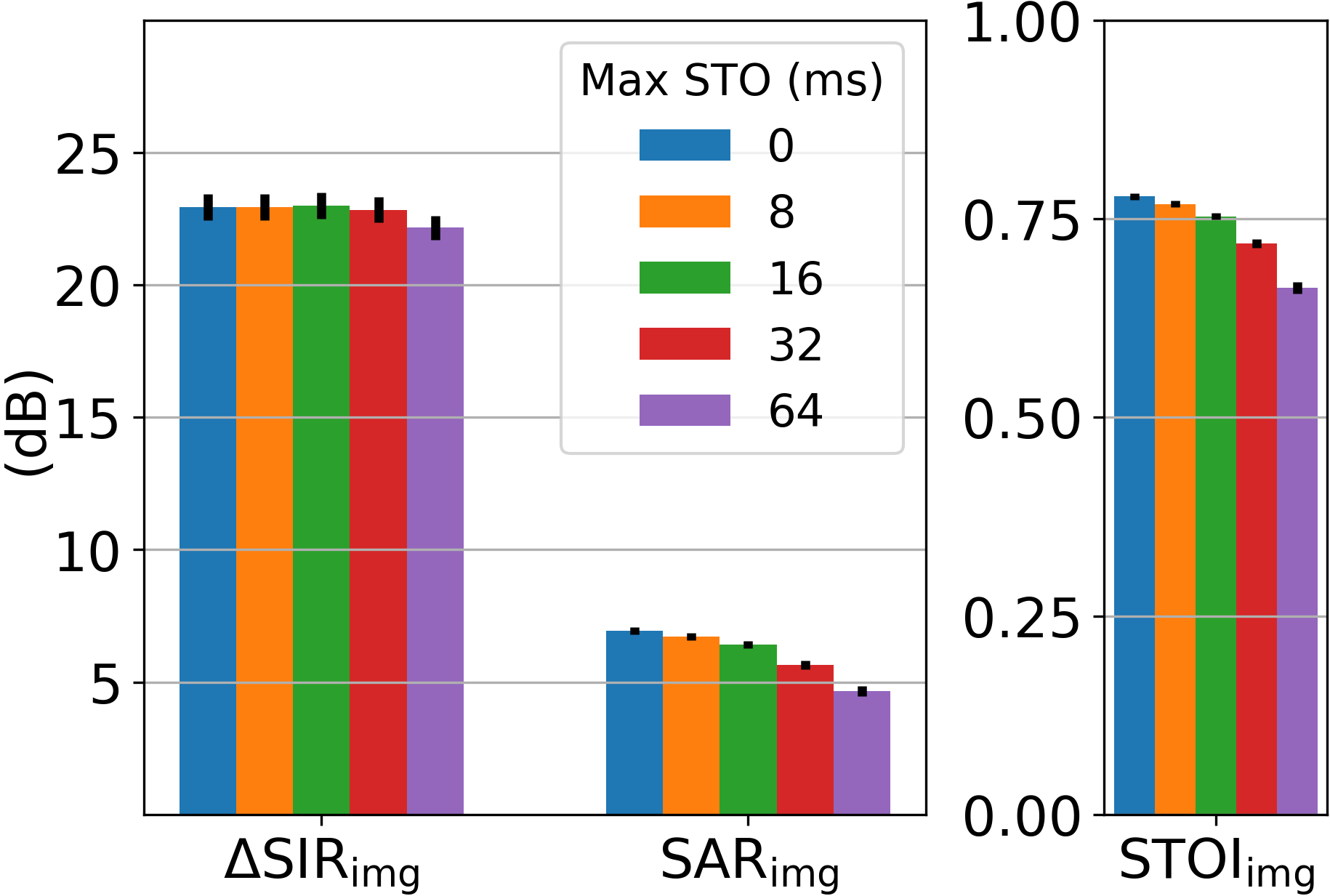}%
		\label{subfig:async_impact_sto}}
	\caption{Impact of \ac{sro} and \ac{sto} on the speech enhancement performance of our system.}
	\label{fig:async_impact}
\end{figure}

The speech enhancement performance of our system under such conditions is reported in Figure~\ref{fig:async_impact} in terms of \acs{sir}\acused{sir}, \acs{sar}\acused{sar}~\cite{Vincent2006} and \acs{stoi}\acused{stoi}~\cite{Taal2010}, where the bars represent the 95~\% confidence interval.
It seems from Figure~\ref{fig:async_impact}\subref{subfig:async_impact_sro} that the \ac{sro} has a limited impact on the performance of our system, even for high values of \acp{sro}. This is probably explained by the fact that the signals are rather short, so that the effect of the \ac{sro} on the signals alignment is limited. Thus, given a rough estimation of the \ac{sro}, resampling the signals with this estimation at large intervals is enough to cope with \acp{sro}.

The impact of \ac{sto} on our system is stronger. The \ac{sir} seems robust to \ac{sto}, probably because of the rank-1 decomposition of the \ac{mwf}. However, the other metrics, especially the \ac{stoi}, are sensitive to this kind of asynchronization, in particular when the \ac{sto} exceeds 16~ms, corresponding to the duration of one frame.

As a conclusion of this section, asynchronization does have a negative impact on our system, especially because of \ac{sto}. In the sequel, we will therefore focus on the impact of \ac{sto} on our distributed speech enhancement system and consider that no \ac{sro} affects the recordings. A solution to compensate for \ac{sto} without resampling the signals is proposed in the next section.

\section{Solution to asynchronization of the input signals of \acp{dnn}}
\label{sec:solution}
We propose to use an attention mechanism to compensate for the negative impact of \ac{sto} on our speech enhancement system. Since the consequence of asynchronization is that the signals recorded on different devices are not aligned in time, we propose to use an alignment mechanism to implicitly shift the asynchronized signals \cite{Bahdanau2014, Luong2015}. To this effect, a temporal alignment attention mechanism is used, which is inspired by the one introduced by Schulze-Forster et al. in a different application field~\cite{Schulze2020}. It is described in the next section.

\subsection{Temporal alignment attention mechanism}
Let $\mathbf{C}_k$ be a reference channel and $\mathbf{C}_j$ an input channel, both of size $T\times F$; let $\mathbf{c}_k(m)$ and $\mathbf{c}_j(n)$ be their $m$-th and $n$-th column respectively. A score between these two columns is computed as:
\[
\tilde{s}_{k, j}(m, n) = \mathbf{c}_k(m) \mathbf{W} {\mathbf{c}_j(n)}^T\,,
\]
where $\cdot^T$ denotes the transpose operator and $\mathbf{W}$ is a learnable matrix. 
In the sequel, we will always consider the first channel of the \acp{mndnn} as the reference channel, so we will drop the index $k$. We have: $\tilde{s}_{j}(m, n) = \tilde{s}_{k, j}(m,n)$.
All the elements $\tilde{s}_{j}(m,n)$ are gathered in the matrix $\tilde{\mathbf{S}}_{j}$ of size $T\times T$. A softmax operation is applied on the rows of $\tilde{\mathbf{S}}_{j}$ to obtain the so-called similarity matrix $\mathbf{S}_{j}$:
\begin{equation}\label{eq:s}
	\mathbf{S}_{j} = \text{softmax}\big(\tilde{\mathbf{S}}_{j}\big)\,.
\end{equation}
The idea of this mechanism is that $\mathbf{S}_{j}$ should contain the probability of the frames of $\mathbf{C}_j$ being aligned with the frames of the reference channel $\mathbf{C}_1$. These probabilities are multiplied with $\{\mathbf{c}_1(i)\}_{i=1..T}$, the columns of $\mathbf{C}_1$ following:
\[
\mathbf{p}_j(m) = \sum_{i=1}^{T}s_{j}(m,i)\mathbf{c}_1(i)\,.
\]
The output matrix $\mathbf{P}_j$ of columns $\{\mathbf{p}_j(m)\}_{m=1..T}$ is concatenated with the input matrix $\mathbf{C}_j$ over the frequency axis. 

\subsection{Integration of the temporal alignment mechanism in Tango}
\begin{figure}
	\includegraphics[width=\linewidth]{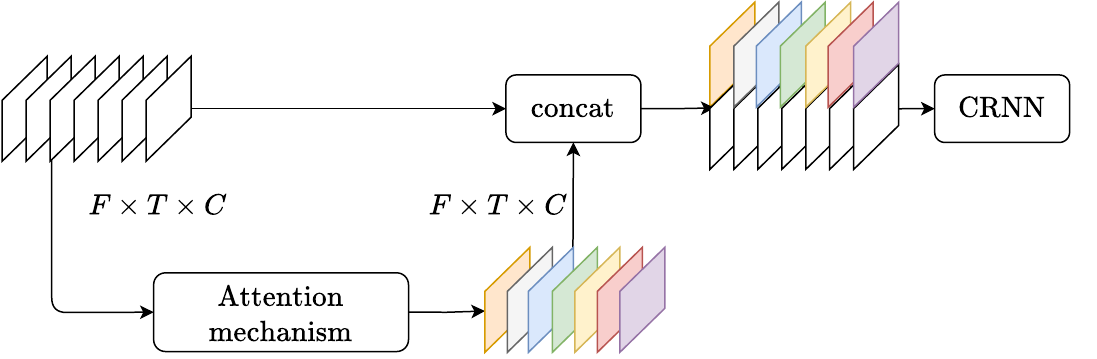}
	\caption{Illustration of the attention-based \ac{crnn}.}
	\label{fig:tamcrnn}
\end{figure}

The previously described attention mechanism is used at the input of the \ac{crnn}. Since only the second filtering step is affected by asynchronization, they are integrated into the \ac{mndnn} only. The new architecture is represented in Figure~\ref{fig:tamcrnn}. Since the input data has twice more features on the frequency axis compared to the initial \ac{mndnn}, the last maximum-pooling layer has a kernel size 8$\times$1 to keep the size of the GRU layer the same. On each node, the first channel, corresponding to the local mixture, is taken as the reference channel. 

\subsection{Quantitative evaluation and analysis}
\begin{figure}
	\includegraphics[width=\linewidth]{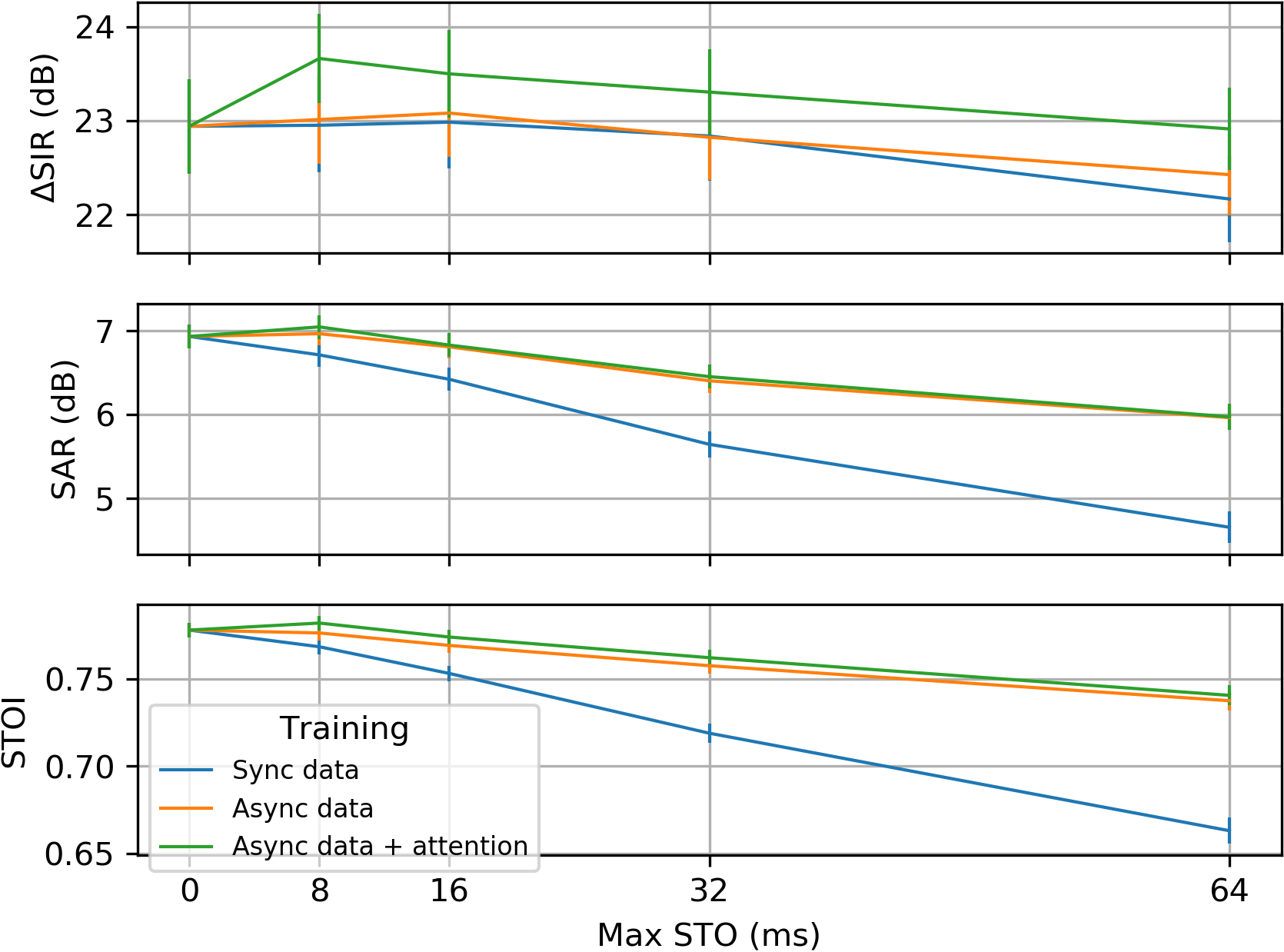}
	\caption{Speech enhancement performance of three different systems.}
	\label{fig:res}
\end{figure}

Three systems are compared to evaluate our solution. The first system is the same as in section~\ref{sec:async}, where the \acp{mndnn} are trained on synchronized data only. The second system has \acp{mndnn} trained on asynchronized data. The third system has \acp{mndnn} with the attention mechanism, trained on asynchronized data. In the asynchronous training set, the \ac{sro} is set to 0~ppm and the \ac{sto} is randomly taken between 0~ms and 32~ms. 
During evaluation, the \ac{sto} is randomly taken between 0~ms and a maximum value $\text{STO}_{\text{max}}$. The same values of $\text{SRO}_{\text{max}}$ as in Section \ref{sec:async} are considered. The results obtained with these three systems are represented in Figure~\ref{fig:res}.
The first conclusion from this experiment is that training the \ac{mndnn} in matching conditions brings robustness to the system in terms of \ac{sar} and \ac{stoi}. However, it does not have any significant impact on the \ac{sir}. With the attention mechanism, even if the differences are not significant, there is a noticeable improvement in terms of \ac{sir} over the system where the \acp{mndnn} are trained in matched conditions but without attention mechanism. This experiment confirms that this attention mechanism is adapted to the misalignment problem, and that it makes our speech enhancement system robust to asynchronization. Another advantage of using such a mechanism is introduced in the next section.

\subsection{Qualitative evaluation and analysis}
\begin{figure}
	\includegraphics[width=\linewidth]{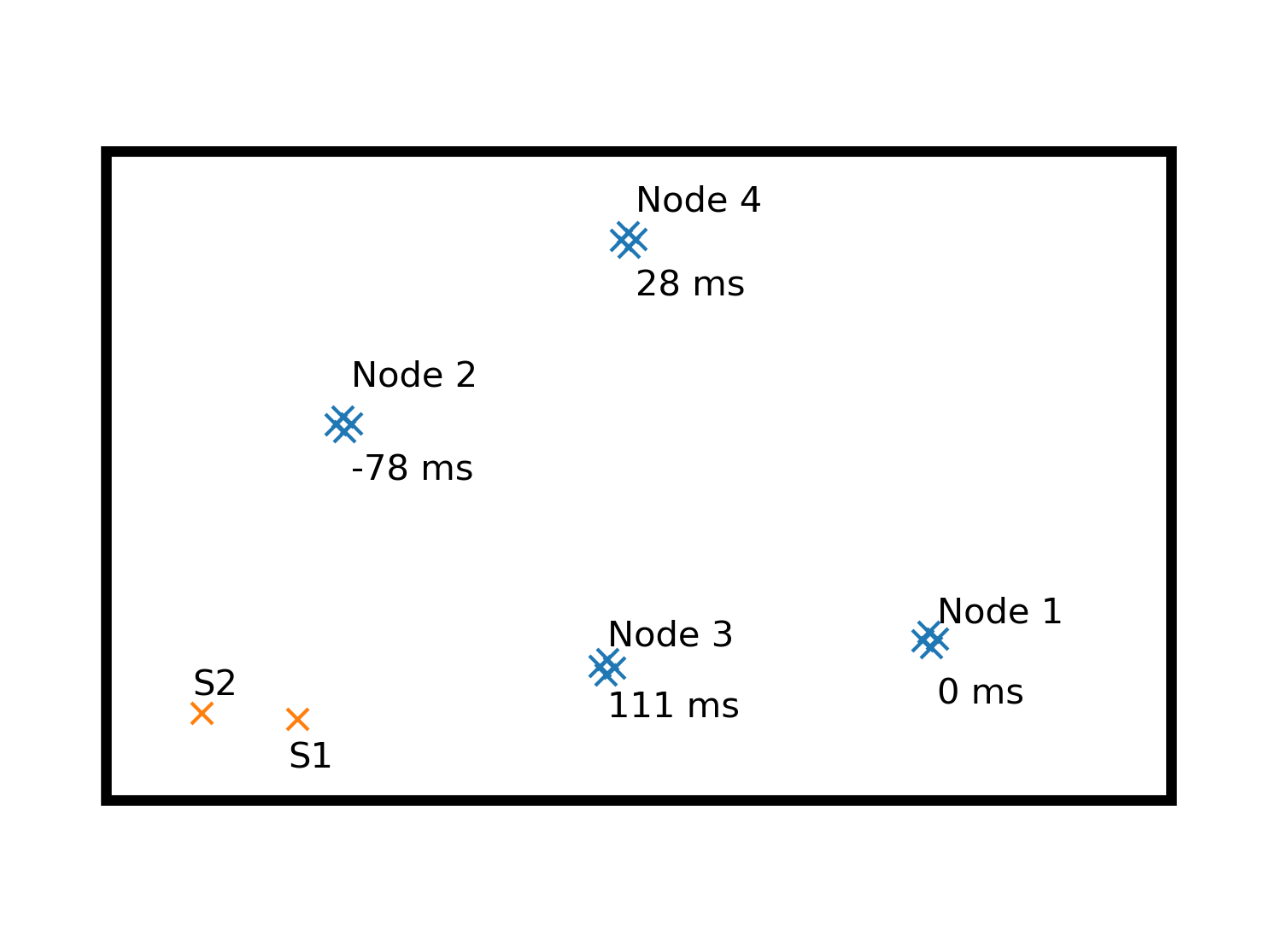}
	\caption{2D view of the evaluation configuration. The \ac{sto} of all nodes, relatively to the first node, it also mentioned.}
	\label{fig:room}
\end{figure}

\begin{figure}
	\includegraphics[width=\linewidth]{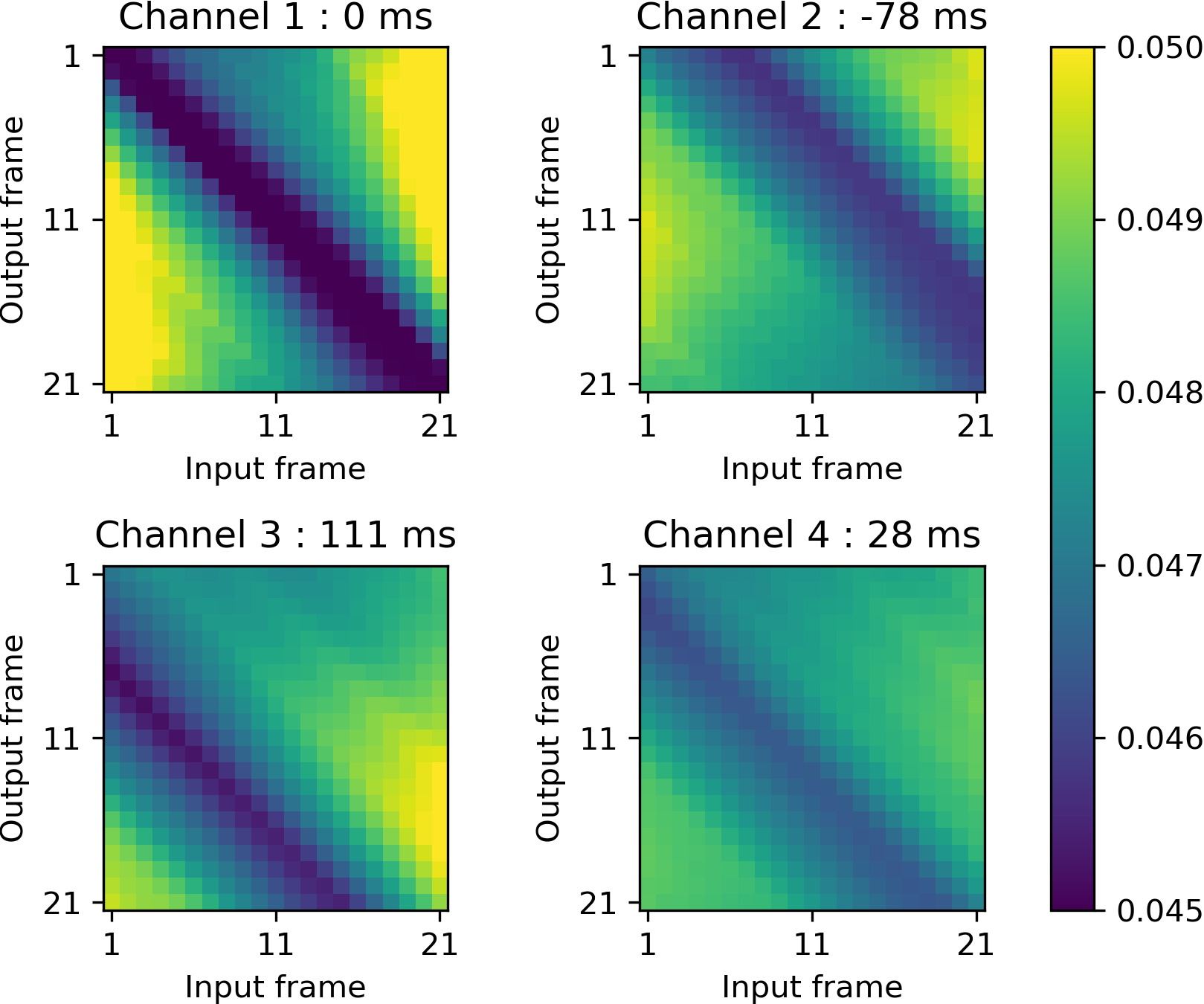}
	\caption{Weights of the similarity matrices $\{\mathbf{S}_{j}\}_{j=1..4}$ (see Eq.~\ref{eq:s}) applied on the four input channels the \ac{mndnn} of the first node of the configuration represented in Figure~\ref{fig:room}.}
	\label{fig:weights}
\end{figure}

In this section, in order to highlight another advantage of using the introduced temporal alignment mechanism, we simulate a specific evaluation room to enhance the behaviour of the attention mechanism. The room configuration is represented in Figure~\ref{fig:room} where the \acp{sto} of all nodes, relatively to the first node, are also mentioned.
We represent in Figure~\ref{fig:weights} the values of the similarity matrices $\{\mathbf{S}_{j}\}_{j=1..4}$ in Equation~\ref{eq:s} computed on the first node of this room configuration. 
These matrices are the weights applied by the first node on the channels at the input of the \ac{mndnn}. It can be seen that these weights seem correlated with the value of the \ac{sto}. For the weights applied on the second channel for example, an upper diagonal can be clearly seen, linking the $i$-th output frames with the $(i+5)$-th input frames. Interestingly enough, the time duration of 5 frames, equal to 80~ms,\footnote{To recall, one frame lasts 16~ms.} corresponds approximatively to the \ac{sto} value of the second channel. Similarly for the weights applied on the third node, the clear diagonal dynamic indicates a correlation between the $i$-th output frames with the $(i-7)$-th input frames, corresponding to a negative delay of approximatively 112~ms, which is almost the \ac{sto} value of the third node relatively to the first node. The same qualitative analysis can be conducted on the similarity matrix applied on the last channel.

As a conclusion of this analysis, the attention mechanism leads to a coarse estimation of the \ac{sto} between asynchronized nodes in an unsupervised manner. This information could be useful to some applications which rely on a rough alignment of the signals \cite{Schmalenstroeer2017}.

\section{Conclusions}
\label{sec:conclusion}
We addressed the issue of asynchronization in a distributed speech enhancement system based on \acp{dnn}. We showed that \ac{sro} had a limited impact on our experiments, but that the influence of \ac{sto} was detrimental to the speech enhancement performance of our system. To cope with it, we introduced a temporal alignment attention mechanism that makes the \acp{dnn} of our system robust to \ac{sto}. In addition, we show that the hidden values of the attention mechanism can be interpreted and that they lead to a coarse estimation of the \ac{sto} at all nodes. We believe that our work introduces a novel and interesting use of attention mechanisms for speech enhancement in ad-hoc microphone arrays. It would be interesting to apply this kind of attention mechanisms on signals in the time domain rather than in the time-frequency domain, where they would probably lead to more precise results and higher performance.

\section{Acknowledgements}

This work was made with the support of the French National Research Agency, in the framework of the  project DiSCogs “Distant speech communication with heterogeneous unconstrained microphone arrays” (ANR-17-CE23-0026-01). Experiments presented in this paper were partially carried out using the Grid5000 testbed, supported by a scientific interest group hosted by Inria and including CNRS, RENATER and several Universities as well as other organizations (see https://www.grid5000).

\bibliographystyle{IEEEtran}
\bibliography{mybib}
\end{document}

%% file: acronyms.tex
\begin{acronym}
\acro{ds}[DSB]{delay-and-sum beamformer}
\acro{mpdr}[MPDR]{minimum power distortionless response beamformer}
\acro{mvdr}[MVDR]{minimum variance distortionless response beamformer}
\acro{lcmp}[LCMP]{linearly constrained minimum power beamformer}
\acro{lcmv}[LCMV]{linearly constrained minimum variance beamformer}
\acro{mwf}[MWF]{multichannel Wiener filter}
\acro{sdw}[SDW-MWF]{speech distortion weighted multichannel Wiener filter}
\acro{mvdr}[MVDR]{minimum variance distortionless response}
\acro{gevd}[GEVD]{generalized eigenvalue decomposition}
\acro{nmf}[NMF-MWF]{non-negative matrix factorization}
\acro{stft}[STFT]{short-time Fourier transform}
\acro{tf}[TF]{time-frequency}
\acro{danse}[DANSE]{distributed adaptive node-specific signal estimation}
\acro{mse}[MSE]{mean squared error}
\acro{wasn}[WASN]{wireless acoustic sensor network}
\acroplural{wasn}[WASNs]{wireless acoustic sensor networks}
\acro{doa}[DOA]{direction of arrival}
\acroplural{doa}[DOAs]{directions of arrival}
\acro{vad}[VAD]{voice activity detector}
\acroplural{vad}[VADs]{voice activity detectors}
\acro{vador}[VADOR]{oracle voice activity detector}
\acroplural{vador}[VADORs]{oracle voice activity detectors}
\acro{irm}[IRM]{ideal ratio mask}
\acroplural{irm}[IRMs]{ideal ratio masks}
\acro{ibm}[IBM]{ideal binary mask}
\acro{dnn}[DNN]{deep neural network}
\acroplural{dnn}[DNNs]{deep neural networks}
\acro{nn}[NN]{neural network}
\acroplural{nn}[NNs]{neural networks}
\acro{lstm}[LSTM]{long short-term memory}
\acro{cnn}[CDNN]{convolutional neural network}
\acroplural{cnn}[CNNs]{convolutional neural networks}
\acro{gru}[GRU]{gated recurrent unit}
\acro{crnn}[CRNN]{convolutional recurrent neural network}
\acroplural{crnn}[CRNNs]{convolutional recurrent neural networks}
\acro{rnn}[RNN]{recurrent neural network}
\acroplural{rnn}[RNNs]{recurrent neural networks}
\acro{tcn}[TCN]{temporal convolutional network}
\acro{dprnn}[DPRNN]{dual-path RNN}
\acroplural{dprnn}[DPRNNs]{dual-path RNNs}
\acro{rir}[RIR]{room impulse response}
\acro{rt}[RT]{reverberation time}
\acroplural{rt}[RTs]{reverberation times}
\acroplural{rir}[RIRs]{room impulse responses}
\acro{ssn}[SSN]{speech shaped noise}
\acro{snr}[SNR]{signal to noise ratio}
\acroplural{snr}[SNRs]{signal to noise ratios}
\acro{sar}[SAR]{source to artifacts ratio}
\acro{sir}[SIR]{source to interferences ratio}
\acroplural{sir}[SIRs]{source to interferences ratios}
\acro{stoi}[STOI]{short term objective intelligibility}
\acroplural{sir}[SIRs]{source to interferences ratios}
\acro{sdr}[SDR]{source to distortion ratio}

\acro{sro}[SRO]{sampling rate offset}
\acro{sto}[STO]{sampling time offset}
\acro{sndnn}[SNDNN]{single-node DNN}
\acro{mndnn}[MNDNN]{multi-node DNN}
\acro{sr}[SR]{sampling rate}
\end{acronym}